\documentclass[a4paper]{article}

\usepackage[english]{babel}
\usepackage[utf8x]{inputenc}
\usepackage[T1]{fontenc}

\usepackage[a4paper,top=3cm,bottom=2cm,left=3cm,right=3cm,marginparwidth=1.75cm]{geometry}
\usepackage{amsmath}
\usepackage{graphicx}
\usepackage[colorinlistoftodos]{todonotes}
\usepackage{amssymb}
\usepackage{bbm}
\usepackage{enumitem}
\usepackage{natbib}
\usepackage{makecell}

\newcommand{\ind}{\perp\!\!\!\!\perp}

\allowdisplaybreaks
\graphicspath{{./figures/}}

\newtheorem{theorem}{Theorem}[section]
\newtheorem{corollary}{Corollary}[theorem]

\title{Clinically Relevant Mediation Analysis using\\Controlled Indirect Effect}

\author{Haoqi Sun, Michael J. Leone, Lin Liu,\\
Shabani S. Mukerji, Gregory K. Robbins, M. Brandon Westover\\
}
\date{}

\begin{document}
\maketitle

\begin{abstract}


Mediation analysis allows one to use observational data to estimate the importance of each potential mediating pathway involved in the causal effect of an exposure on an outcome. However, current approaches to mediation analysis with multiple mediators either involve assumptions not verifiable by experiments, or estimate the effect when mediators are manipulated jointly which precludes the practical design of experiments due to curse of dimensionality, or are difficult to interpret when arbitrary causal dependencies are present. We propose a method for mediation analysis for multiple manipulable mediators with arbitrary causal dependencies. The proposed method is clinically relevant because the decomposition of the total effect does not involve effects under cross-world assumptions and focuses on the effects after manipulating (i.e. treating) one single mediator, which is more relevant in a clinical scenario. We illustrate the approach using simulated data, the ``framing'' dataset from political science, and the HIV-Brain Age dataset from a clinical retrospective cohort study. Our results provide potential guidance for clinical practitioners to make justified choices to manipulate one of the mediators to optimize the outcome.
\end{abstract}

\section{Introduction}

Inferring causal effects and the mediating pathways from observational and/or experimental data is one of the most important problems in healthcare and artificial intelligence~\citep{glass2013causal}. In animal and some human studies, it is possible to conduct a randomized controlled trial (RCT) to infer the causal effect of a particular intervention on an outcome. RCTs are considered the gold standard of causal inference given their ability to limit/reduce multiple sources of bias~\citep{deaton2018understanding}. However, an RCT may not be feasible or ethical for certain interventions. In these cases, researchers must conduct observational studies instead and adjust for potential biases using statistical methods. Advances in statistical methods in causal inference~\citep{pearl2009causality,hernancausal,pingault2018using,yao2020survey} have led to the possibility of studying causal effects in a mathematically principled way using observational data to guide healthcare practice. These methods often allow estimating causal effects in settings where subjects were assigned to an exposure non-randomly based on their characteristics such as age or disease severity at admission~\citep{hernancausal}. Note that throughout this paper we will use the word ``exposure'' instead of ``intervention'' or ``treatment''. ``Exposure'' is more general which includes intervention or treatment, or observational factors such as a disease.
We use the terminology from the potential outcomes framework developed by Neyman, Rubin, and Robins~\citep{neyman1923application,rubin1978bayesian,robins1986new}: when the assignment is equal to the observed exposure, the outcome is called ``factual outcome''; otherwise it is called ``counterfactual outcome''; either of which is called ``potential outcome''.

Mediation analysis is an important sub-field of causal inference. It aims at measuring the relative importance of each mediating pathway, by decomposing the total effect (TE) into parts including mediation due to mediators, and interactions due to the co-existence of exposure and mediators~\citep{vanderweele2015explanation,imai2010general}.
The mediators are defined as those causally affected by the exposure, while also causally affecting the outcome. The categorization of a variable as being a mediator or a confounder is determined by human knowledge or temporal ordering, if any. The total effect can be decomposed in various ways including
\begin{enumerate}[label={(\arabic*)}]
\item controlled direct effect and eliminated effect;
\item natural direct effect and natural indirect effect~\citep{pearl2014interpretation}; and
\item 4-way decomposition: controlled direct effect, reference interaction, mediated interaction, and pure indirect effect~\citep{vanderweele2015explanation}.
\end{enumerate}

The extension of these approaches into multiple mediators with arbitrary causal dependency is challenging:
For decomposition (1), the eliminated effect represents all effects other than the controlled direct effect~\citep{robins2010alternative} which cannot be contributed to each mediator.
For decomposition (2), although the division into natural direct and indirect effects is simple and can be done even in the presence of interaction, this decomposition involves cross-world effects (nested counterfactuals with different exposures), which does not correspond to any randomized experiment performed via interventions on the exposure and/or mediator~\citep{robins2010alternative}; the identification of these effects also requires a strong ``sequential ignorability'' assumption which rules out the possibility of assessing each mediator when they are causally dependent~\citep{tingley2014mediation}. 
For decomposition (3), the pure indirect effect in the case of multiple mediators requires estimating the joint potential outcome, i.e. the potential outcome when the exposure and all mediators would have been set to particular values, which is not clinically practical and suffers from the curse of dimensionality when the number of mediators is large.

In this work, we propose a ``clinically relevant'' mediation analysis approach to decompose the total effect for multiple manipulable mediators with arbitrary causal dependencies, which overcomes the above limitations. Note that we limit the scope to binary exposure and mediator. Here, clinical relevance means
(1) the mediators are manipulable, such as treating a co-morbidity (mediator) of a disease (exposure);
(2) the decomposition involves terms related to the effect due to one mediator being treated, which mimics clinical practice that a physician might focus on treating one mediator at one time, rather than treating all mediators jointly. Intervening on a single mediator makes it possible to rank mediators (comorbidities) based on their CIE (the change in the outcome if everyone's k-th comorbitidy is treated). So that we can make decision to give priority to the top mediators (comorbidities) to spend resources (doctor's attention, medication, research). Since CIE can be viewed as the total effect of mediator on outcome, its effect includes the downstream mediators in case of multiple mediators and confounders; In terms of multiple mediators and confounding, CIE$_k$ includes all its downstream effect, in other words, CIE$_k$ is the total effect of treating the $k$-th mediator on the outcome.; and
(3) there is no cross-world effects, and thus all the quantities in our decomposition can be experimentally validated, such as using the parallel (encouragement) experiment design~\citep{imai2013identification}. In particular, we propose decomposing the total effect into two components for each mediator: the ``controlled direct effect''; and ``scaled controlled indirect effect'' which is a function of CIE (CIE is the effect due to one mediator being treated).



Causal inference and mediation analysis make up an under-represented but scientifically valuable field in artificial intelligence and machine learning applied to healthcare. They help healthcare practitioners and researchers understand the underlying data-generating mechanisms by prospectively or retrospectively observing patients. In general, machine learning algorithms that take causality into account have great potential to guide decision-making in healthcare based not on association but on causality, improving the algorithm performance and transferability to different settings since the causal mechanisms are stable~\citep{peters2017elements}.
The approach developed in this paper provides a new method for mediation analysis that, when applied to a clinical problem, can provide insight into the consequences of preventing or treating a co-morbidity that mediates the effect of a particular disease on a particular outcome. This is a core problem in medicine; much of medicine is devoted to mitigating the effects of a disease by treating a resultant co-morbidity. Our method provides an improved way to quantify the possible effect, or clinical benefit, of such a mitigation strategy on downstream clinical outcomes. Additionally, by allowing arbitrary causal dependencies among multiple mediators, this provides flexibility for a clinician to consider the mediator of interest in clinically realistic scenario.

\section{Related Work}

The existing works mostly focus on the extension of decomposition (2),~i.e. the natural direct and indirect effect approach. \cite{vanderweele2014mediation} and \cite{vanderweele2014effect} extend it to multiple mediators by considering all mediators jointly as one vector-valued mediator, so that the ``sequential ignorability'' assumption (no exposure-induced mediator-outcome confounder) still holds. \cite{daniel2015causal} still estimates the indirect effect of each mediator (although the ``sequential ignorability'' assumption is violated), but uses sensitivity analysis to assess the robustness of their results to the violation of the assumption. As we mentioned above, this approach is not clinically relevant since the natural effects cannot be verified by any experiment. There are also works focusing on the extension of decomposition (3). \cite{bellavia2018decomposition} extends the 4-way decomposition to the finest decomposition that unifies multiple mediators and interactions for causally independent mediators. With more
mediators, it becomes incrementally difficult to define, identify, and estimate these components.

Our approach is closer to the interventional effect approach in~\citep{vanderweele2014effect,vansteelandt2017interventional,loh2019interventional}. The interventional indirect effect is defined as the contrast in the outcome if we fix the exposure, while changing the mediator from a sampled value from the distribution of the mediator among all subjects with one exposure to a sampled value from the distribution from another exposure. However, the sum of interventional direct and indirect effect is not equal to the total effect. In contrast, in our appraoch, the controlled direct effect and the scaled controlled indirect effect add up to the total effect. And we fix the mediator to 0 or 1 for the ease of clinical practice.

\section{Methods}
\label{sec:method}

We use $Y$ to denote the outcome, e.g. mortality, cognitive test score, or a physiological measurement. $A$ denotes the exposure (e.g. taking a pill, infection with HIV or coronavirus, or developing a disease such as Alzheimer's). $M_k$ denotes the $k$-th mediator, e.g. a co-morbid medical condition which worsens the outcome. $L$ denotes the set of covariates, e.g. a patient's age, gender, race, smoking status, and years of education. Here we limit the scope to binary $A$ and $M$; $Y$ is discrete or continuous; and $L$ is a vector of any type of variable. There are $K$ mediators.

\subsection{Total Effect Decomposition}

In general, given a causal DAG, the total effect (TE) can be decomposed into (proof in Appendix~\ref{sec:app_A})
\begin{align}
\label{eq:decomposition_ind}
\text{TE} &= \text{CDE}_k(0) + \text{sCIE}_k \text{ for $k = 1, \ldots, K$} \; ,
\end{align}
where
\begin{align}
\text{CDE}_k(0) &= Y_k(1,0) - Y_k(0,0) \; ;\\
\text{sCIE}_k &= M_k(1)\text{CIE}_k(1) - M_k(0)\text{CIE}_k(0) \; ;\\
\text{CIE}_k(a) &= Y_k(a,1) - Y_k(a,0) \; ;\\
Y_k(a,m) &= Y\left(a,M_1(a),\cdots,M_{ k-1}(a),m,M_{k+1}(a),\cdots,M_K(a)\right) \; ;\\
\label{eq:Ma}
M_k(a) &= M_k\left(a, Pa\{M_k\}(a)\right) \; .
\end{align}

Here we denote $Y(A=a,M_k=m)$, or simply $Y_k(a,m)$, as the potential outcome of $Y$ when $A$ would have been $a$, the $k$-th mediator would have been $m$, and the other mediators were behaving as if A was a. $\text{CDE}_k(0)$ is the controlled direct effect for the $k$-th mediator, defined as the contrast in the potential outcome when the exposure changes from 0 to 1, while fixing the $k$-th mediator to be 0; other mediators were behaving as if A was a. $\text{sCIE}_k$ is the scaled controlled indirect effect for the $k$-th mediator, defined as the controlled indirect effect scaled by the potential outcome of the $k$-th mediator when fixing the exposure to 1, subtracting the same quantity but when fixing the exposure to 0. $\text{CIE}_k(a)$ is the controlled indirect effect of the $k$-th mediator, defined as the contrast in the potential outcome when the $k$-th mediator changes from 0 to 1, while fixing the exposure to $a$ and other mediators were behaving as if A was a. $Pa\{M_k\}(a) = \left\{M_j(a)\right\}_{j\in \text{Parent of }M_k}$ which is the set of causal parents of the $k$-th mediator in the given DAG.

Note that there is no cross-world potential outcome such as $M_k(1, Pa\{M_k\}(0))$. Also note that Equation~\eqref{eq:Ma} is a recursive definition: if there is no parent for $M_k$, it is just $M_k(a)$; if there is a parent mediator $M_1$ of $M_k$, $M_k(a) = M_k(a, M_1(a))$; if $M_2$ is a parent of $M_1$ and $M_1$ is a parent of $M_k$, $M_k(a) = M_k(a, M_1(a), M_2(a, M_1(a)))$; and so forth. If the k-th mediator is not causally affected by the exposure, the $a$ in the parenthesis can be dropped.

We have the following corollary (proof in Appendix~\ref{sec:app_B}):
\begin{corollary}
\label{cor:1}
\begin{equation}
\text{TE} = \frac{1}{K}\sum_{k=1}^K \text{CDE}_k(0) + \frac{1}{K}\sum_{k=1}^K \text{sCIE}_k \; ,
\end{equation}
\end{corollary}
which shows the total effect can also be decomposed as the average of the CDEs of all mediators, and the average of the sCIEs of all mediators, reflecting the average percentage of direct and indirect effects across all mediators. This corollary also provides an alternative way to estimate the total effect, which could serve as a less biased estimate by canceling the model mis-specification biases from each single mediator.
This is a trade of precision for accuracy, because the estimate of the average sCIE is improved, but knowing the contribution of any particular mediator is lost.

\subsection{Interpretation of the Scaled Controlled Indirect Effect}

Suppose (omitting subscript $k$)
\begin{align}
M(1) &= M(0) + \Delta{M} \; ;\\
\text{CIE}(1) &= \text{CIE}(0) + \Delta{C} \; .
\end{align}

We can look at the extreme cases
\begin{align}
\label{eq:sCIE_dC0}
\text{sCIE} &= \Delta M \cdot \text{CIE}(0) = \Delta M \cdot \text{CIE}(1) \qquad\text{if }\Delta C=0 \; ;\\
\label{eq:sCIE_dM0}
\text{sCIE} &= \Delta C \cdot M(0) = \Delta C \cdot M(1) \qquad\text{if }\Delta M=0 \; .
\end{align}
When $\Delta C=0$,~i.e. $\text{CIE}(0)=\text{CIE}(1)$, hence no interaction between the mediator and exposure, sCIE only contains the mediated effect which is the difference in the outcome if that mediator is changed from 0 to 1, scaled by the increase in the probability of the mediator.
When $\Delta M=0$,~i.e. $M(0)=M(1)$, hence no mediation, sCIE only contains the interaction between the mediator and the exposure, scaled by the constant probability of the mediator.
Therefore when $\Delta M\neq 0$ and/or $\Delta C\neq 0$, sCIE is a mixture of mediation and interaction effects. In contrast, CIE is the total effect of mediator on the outcome.

\subsection{Identification Assumptions}
\label{subsec:assumptions}
There are three assumptions needed to identify $M_k(a)$ and $Y_k(a,m)$, and hence $\text{CDE}_k(0)$, $\text{CIE}_k(a)$, and $\text{sCIE}_k$, from observational data. 

1. Consistency assumption: an individual's potential outcome under the observed exposure is equal to the observed outcome
\begin{align}
M_k^{(i)}(a) &= M_k^{(i)} \text{  if } A^{(i)}=a \; ;\\
Y_k^{(i)}(a,m) &= Y^{(i)} \text{  if } A^{(i)}=a, M_k^{(i)}=m \; .
\end{align}
Consistency may be violated if there are multiple versions of exposure~\citep{cole2009consistency}. It is unlikely the case in the ``framing'' dataset. In the case of the HIV-BA dataset, although there are multiple ways to contract HIV-1, we consider HIV-1 infection status as a single exposure because the viral processes in the body following infection are generally similar across patients.

2. Positivity assumption: there is a positive probability of receiving every level of exposure for every combination of values of exposure, mediator of interest, and confounding variables in the population. Usually, large sample size can alleviate this assumption. Positivity assumption is an important assumption for weighting based estimation methods such as inverse propensity weight and doubly robust estimation.

3. Ignorability assumption: the exposed and unexposed subjects have equal distributions of potential outcomes when conditioned on confounding variables. This is sometimes referred as exchangeability assumption.
We need two ignorability assumptions:
\begin{align}
\label{eq:ignorability_M}
M_k(a) \ind A &\,|\, L \; ;\\
\label{eq:ignorability_Y}
Y_k(a,m) \ind A,M_k &\,|\, L \; .
\end{align}
These assumptions can be equivalently expressed as the causal DAG is correct. Hence, we can prove the above equations for multiple mediators with arbitrary causal dependency using d-separation in the single world intervention graph (SWIG)~\citep{richardson2013single}. The proof is given in Appendix~\ref{sec:app_ignorability}. Note that we are not using natural direct or indirect effect, therefore the much stronger sequential ignorability assumption is not needed~\citep{tingley2014mediation}.


\subsection{Effect Estimation}

CDE, CIE and sCIE are defined as functions of $M_k(\cdot)$ and $Y(\cdot)$, which need to be estimated from data. Therefore, the unbiasedness property (consistency, zero bias in the limit of infinite data, not to be confused with the consistency assumption in Section~\ref{subsec:assumptions} for causal inference) partially depends on the unbiasedness of $M_k(\cdot)$ and $Y(\cdot)$ (other than other biases such as selection bias or measurement error).

To this end, we can use doubly robust estimation~\citep{robins1994estimation}, which entails less biased estimation. The doubly robust property is described by a class of models which admits a doubly robust first order influence function~\citep{robins2016technical}. Their influence function has the form of product of two models' influence functions. For example, suppose $Y$ and $A$ are univariate random variables that are dependent on observed data $X$, the expected product of two conditional expectations $\psi(\theta)=\mathbb{E}_\theta [\mathbb{E}_\theta[Y|X]\cdot\mathbb{E}_\theta[A|X]]$ is a doubly robust estimator~\citep{cui2019selective}; the other well-known example is the doubly robust estimator for the total effect (average treatment effect), which is unbiased if at least one of the outcome ($f$ function below) or propensity model ($g$ function below) is unbiased.


The doubly robust estimator is written as
\begin{align}
\label{eq:dre_M}
M_k(a) &\approx \frac{1}{N}\sum_{i=1}^N \left[ f_{M,k}^{(i)} + \frac{\mathbbm{1}(A^{(i)}=a)}{g_M^{(i)}}\left(M_k^{(i)}-f_{M,k}^{(i)}\right)\right]\; ;\\
\label{eq:dre_Y}
Y_k(a,m) &\approx \frac{1}{N}\sum_{i=1}^N \left[ f_{Y,k}^{(i)} + \frac{\mathbbm{1}(A^{(i)}=a,M_k^{(i)}=m)}{g_{Y,k}^{(i)}}\left(Y^{(i)}-f_{Y,k}^{(i)}\right)\right] \; .
\end{align}
where
\begin{align}
f_{M,k}^{(i)} &= \mathbb{E}\left[M_k | A=a,L^{(i)}\right] \; ;\\
f_{Y,k}^{(i)} &= \mathbb{E}\left[Y | A=a,M_k=m,L^{(i)}\right] \; ;\\
g_M^{(i)} &= P(A^{(i)}|L^{(i)}) \; ;\\
g_{Y,k}^{(i)} &= P(M_k^{(i)}|L^{(i)},A^{(i)})P(A^{(i)}|L^{(i)}) \; .
\end{align}


\subsection{Model Selection and Fitting}
\label{subsec:perturbation}

Here we used the principled approach introduced in~\citep{cui2019selective}. In estimating either TE of exposure on outcome, or CIE (TE of of mediator on outcome), we want to minimize the bias $\mathbb{E}[\psi'-\psi]$, where $\psi$ is the ground truth TE and $\psi'$ is the estimated TE. Directly minimizing this bias is impossible due to unknown $\psi$. Instead, we minimize a pseudo-risk over different choice of models, where the optimal model choice is least sensitive to perturbations due to model mis-specification. Here we used the mixed minmax solution, which is proved to have a doubly robust property, i.e.~zero bias if at least one candidate estimation model is correctly specified. Here, we choose from (1) $\ell_2$-norm penalized linear regression or logistic regression; (2) $\ell_2$-norm penalized support vector machine (SVM) classifier; (3) random forest; and (4) XGBoost, a type of gradient boosting tree. For the ordinal outcome in the framing dataset introduced later, we used pairwise approach~\citep{liu2009learning} to convert ordinal regression problem into binary classification and then solved using the above models.

We used nested cross-validation to fit the models. Nested cross-validation consists of an inner loop and an outer loop. The purpose of the outer loop is to compute an unbiased estimate when applied to data not part of the training set. The purpose of the inner loop was to find the best hyper-parameter, $C$ the strength of $\ell_2$-norm penalty, to avoid overfitting. The outer loop divided the data into multiple folds. Each fold was used as the testing set, while the other folds were combined and further divided into inner folds. Each inner fold was used as the validation set, while the other inner folds were combined as the training set. The model was trained with a particular $C$ on the training set and evaluated on the validation set. The $C$ with the best average validation performance was chosen and re-fit using the combined training and validation sets. The model was then used to estimate the causal effects on the testing set. The final reported effects were the average effects on the testing sets from the outer loop. The confidence intervals were obtained using bootstrapping 1,000 times.

\section{Results}
\label{sec:result}

\subsection{Dataset}



The simulated data is generated based on the causal ordering implicated by the DAG, i.e.~$L \longrightarrow A \longrightarrow M \longrightarrow Y$. Each variable is generated as a generalized linear function of its causal parents plus noise. We first randomly generate the coefficients, take the inner product between the coefficients and causal parents plus intercept. The intercept is manually chosen to make the average of the inner product zero. We then added Gaussian noise with standard deviation 1. For binary variables such as $A$ and $M$, we further applied the sigmoid transformation, and binarized it using a threshold of 0.5. The sample size $N$ is 1,000; the number of covariates in $L$ is 2; and the number of mediators in $M$ is 2 or 3 depending on the DAG we study.

We also used a public dataset ``framing'' used in the R package ``mediation''~\citep{tingley2014mediation}. The detailed description of the framing data can be found in~\citep{brader2008triggers}. It is a randomized experiment in which the subjects are shown immigration stories with different framing. The exposure is whether the story is framed positively and features an European immigrant. The covariates include age, gender, education level, and income. The mediators include negative emotion and perceived harm. Emotion measures subjects' negative feeling, and is converted to 1 if more or equal to 8. Perceived harm is with respect to increased immigration, and is converted to 1 if more or equal to 7. The outcome is a four-point scale measuring the attitudes toward increased immigration. There are 265 subjects in this dataset. Note that since the exposure is randomized, we used outcome regression instead of doubly robust estimation for this dataset.

The ``HIV-Brain Age'' (HIV-BA) dataset comes from a retrospective cohort study
which investigates the effect of HIV-1 infection (exposure) on brain age index (BAI) predicted by the sleep electroencephalogram (EEG)~\citep{anonymous} (outcome) through multiple mediating comorbidities and side-effects. The outcome BAI is in unit of years, and bigger value represents older age, hence worse outcome. The cohort is composed of participants with a possible sleep disorder who underwent a full-night diagnostic sleep study at a hospital's
sleep lab. The HIV+ subset were those who were diagnosed with HIV infection prior to their sleep study and are currently on antiretroviral therapy based on clinical chart review. The HIV- subset never had HIV infection. The exposure is HIV infection (binary). The covariates are age, gender, race, alcoholism and smoking history. The mediators are hyperlipidemia, heart valve disorders, and insomnia (all binary). The outcome is the brain age index, which is a continuous number in unit of years. There are 43 HIV+ and 3,048 HIV- subjects.

\subsection{Causally Independent Mediators}

\begin{figure}[h]
\includegraphics[width=0.4\textwidth]{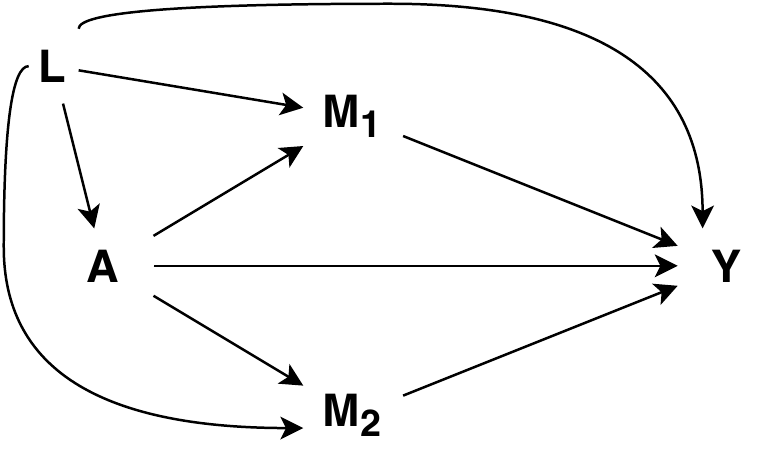}
\centering
\caption{Causal graph with independent mediators. The arrows represent causal influences. $M_1$ and $M_2$ are the first and second mediators respectively.}
\label{fig:causal_graph_ind}
\end{figure}

We assume two causally independent mediators as shown in \figurename~\ref{fig:causal_graph_ind}. It represents the case that $A$ takes effect on $Y$ through two independent mechanisms $M_1$ and $M_2$. Note that here we use the example of 2 mediators, but in general it can be multiple.

\subsubsection{Simulation experiments}
\label{sec:sim_data_ind}

The results are shown in Table~\ref{tab:indep_simulated_data}. The model selection method described in Section~\ref{subsec:perturbation} correctly selected linear models (logistic regression and linear regression) for the propensity models and outcome models when estimating $M_k(a)$ and $Y_k(a,m)$. By "correct" we mean that the data is generated using a generalized linear model. Since this is simulated data, we can get the ground truth effects by directly manipulating $A$ and $M$'s. All effects except CDE for $M_1$ and $M_2$ and the total effect for $M_1$ is within the 95\% confidence interval. The bias in estimating CDE, and hence in total effect, is due to the bias in the estimated coefficient when using the $\ell_2$ penalized linear models. The confidence interval for sCIE is in general wider than that for CIE because sCIE is a function of CIE$(a)$ and $M(a)$ which jointly considers the exposure and mediator.

\begin{table}[htbp]
\caption{The estimated effects and their ground truth value for the simulated data with independent mediators}
\begin{center}
\begin{tabular}{l|cc|cc}
\hline
Effect & $M_1$ & \makecell{$M_1$ True Effect \\ from Simulation} & $M_2$ & \makecell{$M_2$ True Effect \\ from Simulation} \\ \hline
\%CDE & 78.5 [67.2 -- 90.3] & 93.9 & 72.4 [61.7 -- 82.9] & 87.4 \\ \hline
\%sCIE & 21.5 [9.7 -- 32.8] & 6.1 & 27.6 [18.0 -- 38.3] & 12.6 \\ \hline
CDE & 0.99 [0.75 -- 1.24] & 1.15 & 0.95 [0.73 -- 1.17] & 1.07 \\ \hline
sCIE & 0.27 [0.12 -- 0.42] & 0.075 & 0.36 [0.22 -- 0.53] & 0.16 \\ \hline
TE & 1.27 [1.00 -- 1.53] & 1.23 & 1.31 [1.09 -- 1.60] & 1.23 \\ \hline
CIE0 & 4.04 [3.72 -- 4.35] & 4 & 4.71 [4.44 -- 5.00] & 5 \\ \hline
CIE1 & 4.10 [3.78 -- 4.39] & 4 & 4.84 [4.56 -- 5.11] & 5 \\ \hline
\end{tabular}
\end{center}
\label{tab:indep_simulated_data}
\end{table}

\subsubsection{Framing dataset}
\label{sec:framing_data_ind}

For the framing dataset, we used emotion and perceived harm as the two independent mediators. The model selection method selected linear SVM for the outcome models when estimating $Y_k(a,m)$ (outcome regression is used since the exposure is assigned at random). The result is shown in Table~\ref{tab:indep_real_data}, which is consistent with the finding that emotion (35.6\% sCIE) is a leading mediator compared to perceived harm (18.8\% sCIE) when people are making decisions about immigration. But interestingly, the CIE of perceived harm is higher than emotion. In other words, directly reducing perceived harm could be more effective than directly improving the negative emotion (directly intervene the mediator), but it is more difficult to induce perceived harm than to induce negative emotion using different ways of framing (change mediator by intervening the exposure), due to the scaling of mediation effect as well as interaction effect (Equation~\eqref{eq:sCIE_dC0} and~\eqref{eq:sCIE_dM0}).
The total effect estimated in \citep{tingley2014mediation} Section 6.2 is 0.42 (95\% confidence interval [0.17--0.62]). Our estimation is 0.31 to 0.36.

\begin{table}[h]
\caption{The estimated effects for the framing dataset when assuming independent mediators}
\begin{center}
\begin{tabular}{cll}
\hline
Effect & \multicolumn{1}{c}{Emotion ($M_1$)} & \multicolumn{1}{c}{Perceived Harm ($M_2$)} \\ \hline
\%CDE & 43.4 [-119.6 -- 105.1] & 61.0 [-76.0 -- 186.0] \\ \hline
\%sCIE & 56.6 [-5.1 -- 219.6] & 39.0 [-86.0 -- 176.0] \\ \hline
CDE & 0.16 [-0.12 -- 0.77] & 0.19 [-0.12 -- 0.86] \\ \hline
sCIE & 0.20 [-0.014 -- 0.29] & 0.12 [-0.13 -- 0.20] \\ \hline
TE & 0.36 [0.026 -- 0.81] & 0.31 [-0.038 -- 0.79] \\ \hline
CIE0 & 0.74 [0.41 -- 1.22] & 1.03 [0.58 -- 1.43] \\ \hline
CIE1 & 0.79 [0.38 -- 1.10] & 1.05 [0.51 -- 1.29] \\ \hline
\end{tabular}
\end{center}
\label{tab:indep_real_data}
\end{table}


\subsection{Causally Dependent Mediators}

\begin{figure}[h]
\includegraphics[width=0.4\textwidth]{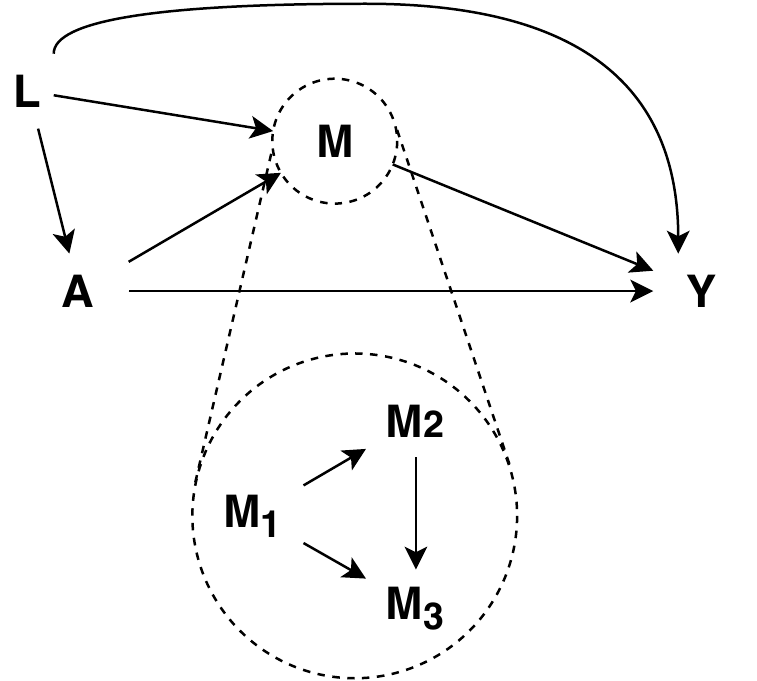}
\centering
\caption{Causal graph with dependent mediators. The arrows represent causal influences. Here we use an example of 3 mediators. $M_1$, $M_2$, and $M_3$ are the mediators which are causally affected as in the figure. Both $L$ and $A$ causally affect each mediator; each mediator causally affect the outcome $Y$.}
\label{fig:m1tom2}
\end{figure}

We assume three causally dependent mediators as shown in \figurename~\ref{fig:m1tom2}. It represents the case that $A$ has an effect on $Y$ through three mechanisms $M_1$, $M_2$, and $M_3$, while $M_1$ also causes $M_2$ and $M_3$, and $M_2$ also causes $M_3$. Note that here we use the example of 3 mediators, but in general it can be multiple and arbitrary causal dependencies as long as there are no cycles.

\subsubsection{Simulation experiments}
\label{sec:sim_data_dep}

In Table~\ref{tab:dep_simulated_data} we show the result. The model selection method described in Section~\ref{subsec:perturbation} again correctly selected linear models (logistic regression and linear regression) for the propensity models and outcome models when estimating $M_k(a)$ and $Y_k(a,m)$. Since this is simulated data, we can get the ground truth effects by directly manipulating $A$ and $M$'s. The true effects are within the 95\% confidence interval for $M_1$ and $M_3$. $M_2$ tends to overestimate the indirect effect and underestimate the direct effect.

\begin{table}[htbp]
\caption{The estimated effects and their ground truth value for the simulated data with dependent mediators}
\begin{center}
\begin{tabular}{l|cc|cc|cc}
\hline
Effect & $M_1$ & \makecell{$M_1$ True\\Effect} & $M_2$ & \makecell{$M_2$ True\\Effect} & $M_3$ & \makecell{$M_3$ True\\Effect} \\ \hline
\%CDE & 89.057 [79.7 -- 97.9] & 92.5 & 85.6 [57.0 -- 100.1] & 86.5 & 90.9 [82.0 -- 103.1] & 92.4 \\ \hline
\%sCIE & 10.9 [2.1 -- 20.3] & 7.5 & 14.4 [-0.13 -- 43.0] & 13.5 & 9.1 [-3.12 -- 18.0] & 7.6 \\ \hline
CDE & 1.17 [0.87 -- 1.43] & 1.26 & 1.24 [0.44 -- 1.59] & 1.15 & 1.24 [0.89 -- 1.70] & 1.21 \\ \hline
sCIE & 0.14 [0.028 -- 0.26] & 0.10 & 0.21 [0.001 -- 0.64] & 0.18 & 0.13 [-0.041 -- 0.24] & 0.1 \\ \hline
TotalEffect & 1.31 [1.04 -- 1.59] & 1.36 & 1.45 [0.74 -- 1.79] & 1.33 & 1.37 [1.02 -- 1.71] & 1.31 \\ \hline
CIE0 & 2.82 [2.56 -- 3.09] & 3.75 & 1.77 [1.27 -- 2.19] & 2.18 & 1.16 [0.84 -- 1.50] & 1 \\ \hline
CIE1 & 2.83 [2.57 -- 3.10] & 3.81 & 1.77 [1.40 -- 2.46] & 2.2 & 1.06 [0.66 -- 1.40] & 1 \\ \hline
\end{tabular}
\end{center}
\label{tab:dep_simulated_data}
\end{table}

\subsubsection{HIV-Brain age dataset}
\label{sec:hiv_ba_dep}

People with HIV take antiretroviral therapy drugs, where some of the drugs, such as Lopinavir, Saquinavir, and Stavudine is associated with high cholesterol level in the blood (hyperlipidemia)~\citep{feeney2011hiv}, which in turns increases the risk of heart disorders such as heart valve disorder~\citep{kleinauskiene2018degenerative}, and eventually leads to sleep disorders such as insomnia. Therefore, in this case $M_1$ is hyperlipidemia; $M_2$ is heart valve disorder; and $M_3$ is the insomnia. The model selection method selected linear models (logistic regression for exposure and linear regression for outcome) when estimating $M_k(a)$ and $Y_k(a,m)$.

\begin{table}[htbp]
\caption{The estimated effects (in years of brain age) for the HIV-BA dataset when assuming multiple dependent mediating comorbidities}
\begin{center}
\begin{tabular}{llll}
\hline
Effect & \multicolumn{1}{c}{Hyperlipidemia ($M_1$)} & \multicolumn{1}{c}{Heart valve disorder ($M_2$)} & \multicolumn{1}{c}{Insomnia ($M_3$)} \\ \hline
\%CDE & 13.7 [-73.8 -- 124.3] & 44.3 [-48.3 -- 134.3] & 89.8 [-29.6 -- 413.5] \\ \hline
\%sCIE & 86.3 [-24.3 -- 173.8] & 55.7 [-34.3 -- 148.3] & 10.2 [-313.5 -- 129.6] \\ \hline
CDE & 0.69 [-1.16 -- 3.19] & 2.07 [-1.16 -- 5.20] & 3.24 [-0.018 -- 6.22] \\ \hline
sCIE & 4.36 [-0.55 -- 8.22] & 2.60 [-0.82 -- 7.00] & 0.37 [-3.07 -- 7.01] \\ \hline
TE & 5.06 [0.45 -- 8.58] & 4.67 [0.84 -- 8.49] & 3.60 [0.081 -- 8.97] \\ \hline
CIE0 & -0.87 [-1.84 -- -0.002] & 2.05 [0.86 -- 3.19] & 0.73 [-0.15 -- 1.60] \\ \hline
CIE1 & 5.35 [-1.21 -- 9.84] & 7.22 [-1.16 -- 14.4] & 0.84 [-5.05 -- 11.12] \\ \hline
\end{tabular}
\end{center}
\label{tab:dep_real_data2}
\end{table}

The results indicate that hyperlipidemia is an important comorbidity in HIV+ subjects. The relatively large CIE means that directly treating hyperlipidemia has an substantial effect on brain age index in HIV+ subjects; the relatively large sCIE means that HIV infection itself can substantially increases the prevalence of hyperlipidemia, which subsequently has both high mediation and interaction effects on brain age index. Heart valve disorder also has relatively large CIE; but smaller sCIE, indicating the relatively weaker increase in the prevalence of heart valve disorder due to HIV infection and hyperlipidemia and the interaction with them. On the other hand, insomnia has limited effect on brain age index compared to hyperlipidemia and heart valve disorder. Due to the limited number of 43 HIV+ subjects, the confidence interval is very wide for HIV+ subjects, indicating the importance of having enough samples for mediation analysis (detailed in Limitations in Section~\ref{sec:discussion}).



\section{Discussion}
\label{sec:discussion}

We have presented a clinically relevant method of mediation analysis with multiple manipulable mediators and arbitrary causal dependency, using observational data. Our approach is clinically relevant because it makes the observational data useful for doctors to think about clinical decision-making, as detailed in the following aspects: Since the decomposition eliminates cross-world considerations, the effects are directly related to what would happen if they took a particular course of action to treat one comorbidity. The elimination of cross-world considerations is also a lead-in to confirmation of a hypothesis in a clinical trial. It is also clinically practical since the controlled indirect effect focuses on the effect of manipulating (treating) one single mediator rather than all of them jointly.

\paragraph{Alternative interpretations of the scaled controlled indirect effect} We have
\begin{align}
\text{sCIE} &= (M(0) + \Delta{M})(\text{CIE}(0) + \Delta{C}) - M(0)\text{CIE}(0) \notag\\
&= \Delta{C} \cdot M(0) + \Delta{M} \cdot \text{CIE}(0) + \Delta{M} \cdot \Delta{C} \notag\\
&= \Delta{C} \cdot M(0) + \Delta{M} \cdot \text{CIE}(1) \; .
\end{align}

The last equation shows the consequences of reducing $\Delta M$.
The meaning of reducing $\Delta M$ is intuitive. If a certain medication or preventive measure reduces the risk of the mediator by a known percent, that percent multiplied by CIE(1) is the amount of outcome prevented, averaged across the exposed population.
On the other hand, sCIE can also be viewed as the effect of the mediator of interest on the outcome in the exposed population ($\Delta M \cdot \text{CIE}(1)$) beyond the baseline level of exposure-mediator interaction in the unexposed population ($\Delta C \cdot M(0)$).

\paragraph{Cross validation and model estimation in causal inference} Regularization and using cross validation to select the regularization strength is in general not advised in effect estimation, since the loss function of regularized models do not respect the target causal effect. The idea of using perturbation as a pseudo-risk, as used in Section~\ref{subsec:perturbation}, represents a possible direction. Other possibilities include optimizing regularization that improves the consistency assumption, such as minimizing the difference between the factual branch of model-based potential outcome vs. the observed value, such as in \cite{parikh2018malts}.

\paragraph{Extension to path-specific analysis} Path-specific analysis is an extension to mediation analysis by looking at the effect mediated by a path (a bundle of nodes and edges)~\citep{shpitser2013counterfactual}. Longitudinal setting represents a typical use case in path-specific analysis~\citep{nabi2018estimation}. The idea is TE$ = (Y(a)-Y_\pi) + (Y_\pi-Y(a'))$, where $Y_\pi$ is the effect specific to path pi. The decomposition is analogous to TE=$[Y(a,M(a))-Y(a,M(a'))] + [Y(a,M(a')-Y(a',M(a'))]=$NIE+NDE, which still requires cross-world counterfactuals. In contrast, our approach represents a ``controlled'' flavor, vs. ``natural'' flavor, which is TE = CDE+sCIE = CDE(0)+f(CIE(1), CIE(0)), and may be extended to TE $= [Y(a, do(M_\pi=0)) - Y(a', do(M_\pi=0))] + f[Y(a, do(M_\pi=m)) - Y(a, do(M_\pi=m')), Y(a', do(M_\pi=m)) - Y(a', do(M_\pi=m'))]$. The controlled flavor has the advantages that CIE directly simulates what if the mediator path are intervened, answering the clinically relevant question: what if I intervene the mediating path? The disadvantage is that CIE is TE of the mediator on outcome, which is subject to unmeasured confounding. The natural flavor has the advantage that it deals with unmeasured confounding, but NIE and NDE cannot be interpreted in the clinical relevant way. As an important future work, extension of the controlled flavor into path-specific effects is needed.

\paragraph{Limitations} First, our analysis is limited to the case where both $A$ and $M$ are binary (0 or 1) making it restrictive in applications. Although it is a helpful simplification to indicate if the mediator (comorbidity) is treated or not, in reality comorbidities can be reduced without being fully treated.
Second, we have not considered other types of contrast. In the present work we have focused on the difference between two potential outcomes. But depending on the data type of $Y$ and $M$, different decomposition equations need to be derived and validated.

The analysis of HIV-BA dataset is limited in terms of the number of HIV+ patients. Mediation analysis requires a relatively large sample size. This is because mediation analysis divides the data into multiple strata,~i.e. samples with and without the presence of each mediator in both exposed and unexposed groups. And there should be enough samples in each stratum to reduce sampling bias. In the case of nested cross-validation, the sample size should be even larger to make sure each fold in the inner loop has enough samples. The fact that our approach deals with each mediator one by one reduces the need for large sample size so that the samples need not grow with the number of mediators. This is helpful but does not completely resolve this limitation. Monte Carlo based power analysis can be done by generating the data using models estimated from actual data, up to the point significance is shown~\citep{schoemann2017determining}.

Other limitations are, as in all causal inference studies, we did not consider all potential biases in the real data examples, including
(1) unmeasured confounding, i.e.~incomplete or incorrect variable list in $L$. We have not done sensitivity analysis to address this;
(2) selection bias, especially in the HIV-BA dataset, the dataset comes from a hospital sleep lab, where the prevalence of sleep disorders is higher than that in the general population; and
(3) measurement noise, i.e.~possible subjectivity in the framing dataset, and measurement noise in predicted brain age in the HIV-BA dataset since it is based on a single night of brain activity monitoring, not multiple.

\section{Conclusion}
\label{sec:conclusion}

The proposed approach can be used to assess the importance of multiple manipulable mediators with arbitrary causal dependencies. In the case of healthcare problems where the mediators are comorbidities or side-effects of certain exposures, our approach provides principled guidance for choosing which mediator to treat in order to optimize the healthcare outcome.

\section*{Acknowledgement}
This work was supported by the Developmental Award from the Harvard University Center for AIDS Research (HU CFAR NIH/NIAID fund 5P30AI060354-16).

\bibliographystyle{ieeetr}
\bibliography{main}

\newpage
\appendix
\section*{Supplemental Material}

\section{Proof of Equation~\eqref{eq:decomposition_ind}}
\label{sec:app_A}

We have the total effect as
\begin{align}
\label{eq:app1}
\text{TE} &= Y(a=1) - Y(a=0) \notag\\
&= Y(1) - Y(0) \notag\\
&= Y\left(1,M_1(1,Pa\{M_1\}(1)),\cdots,M_K(1,Pa\{M_K\}(1))\right) \notag\\
&\qquad- Y\left(0,M_1(0,Pa\{M_1\}(0)),\cdots,M_K(0,Pa\{M_K\}(0))\right) \; .
\end{align}

Expanding the $k$-th mediator, we have
\begin{align}
Y(1) &= Y\left(1,M_1(1,Pa\{M_1\}(1)),\cdots,M_K(1,Pa\{M_K\}(1))\right) \notag\\
&= Y\left(1,M_1(1,Pa\{M_1\}(1)),\cdots,1,\cdots,M_K(1,Pa\{M_K\}(1))\right)M_k(1,Pa\{M_k\}(1)) \notag\\
&\qquad+ Y\left(1,M_1(1,Pa\{M_1\}(1)),\cdots,0,\cdots,M_K(1,Pa\{M_K\}(1))\right)\left(1-M_k(1,Pa\{M_k\}(1))\right) \notag\\
&= Y\left(1,M_1(1,Pa\{M_1\}(1)),\cdots,0,\cdots,M_K(1,Pa\{M_K\}(1))\right) \notag\\
&\qquad+ M_k(1,Pa\{M_k\}(1))\Big[Y\left(1,M_1(1,Pa\{M_1\}(1)),\cdots,1,\cdots,M_K(1,Pa\{M_K\}(1))\right) \notag\\
&\qquad\qquad\qquad\qquad\qquad\qquad- Y\left(1,M_1(1,Pa\{M_1\}(1)),\cdots,0,\cdots,M_K(1,Pa\{M_K\}(1))\right)\Big] \notag\\
\label{eq:y1_expand_k}
&= Y\left(1,M_1(1,Pa\{M_1\}(1)),\cdots,0,\cdots,M_K(1,Pa\{M_K\}(1))\right) + M_k(1,Pa\{M_k\}(1))\text{CIE}_k(1) \; .
\end{align}

Similarly, we have
\begin{align}
Y(0) &= Y\left(0,M_1(0,Pa\{M_1\}(0)),\cdots,M_K(0,Pa\{M_K\}(0))\right) \notag\\
\label{eq:y0_expand_k}
&= Y\left(0,M_1(0,Pa\{M_1\}(0)),\cdots,0,\cdots,M_K(0,Pa\{M_K\}(0))\right) + M_k(0,Pa\{M_k\}(0))\text{CIE}_k(0) \; .
\end{align}

Therefore,
\begin{align}
TE &= Y(1)-Y(0) \notag\\
&= \Big[Y\left(1,M_1(1,Pa\{M_1\}(1)),\cdots,0,\cdots,M_K(1,Pa\{M_K\}(1))\right) \notag\\
&\qquad- Y\left(0,M_1(0,Pa\{M_1\}(0)),\cdots,0,\cdots,M_K(0,Pa\{M_K\}(0))\right)\Big] \notag\\
&\qquad+ \Big[M_k(1,Pa\{M_k\}(1))\text{CIE}_k(1) - M_k(0,Pa\{M_k\}(0))\text{CIE}_k(0)\Big] \notag\\
&= \underbrace{\Big[Y_k\left(1,0\right) -  Y_k\left(0,0\right)\Big]}_{\text{CDE}_k(0)} + \underbrace{\Big[M_k(1)\text{CIE}_k(1) - M_k(0)\text{CIE}_k(0)\Big]}_{\text{sCIE}_k} \; .
\end{align}

\section{Proof of Corollary~\ref{cor:1}}
\label{sec:app_B}

Equation~\eqref{eq:y1_expand_k} and~\eqref{eq:y0_expand_k} are general equations obtained by expanding the $k$-th mediator. We repeat this for all mediators 1, \dots, $K$, so that
\begin{align}
TE &= \text{CDE}_1(0) + \text{sCIE}_1 \; ;\\
&\cdots \notag\\
TE &= \text{CDE}_K(0) + \text{sCIE}_K \; .
\end{align}
Therefore,
\begin{align}
TE &= \frac{1}{K}\sum_{k=1}^K \text{CDE}_k(0) + \frac{1}{K}\sum_{k=1}^K \text{sCIE}_k \; .
\end{align}

\section{Proof of Ignorability}
\label{sec:app_ignorability}

\begin{figure}[h]
\includegraphics[width=0.9\textwidth]{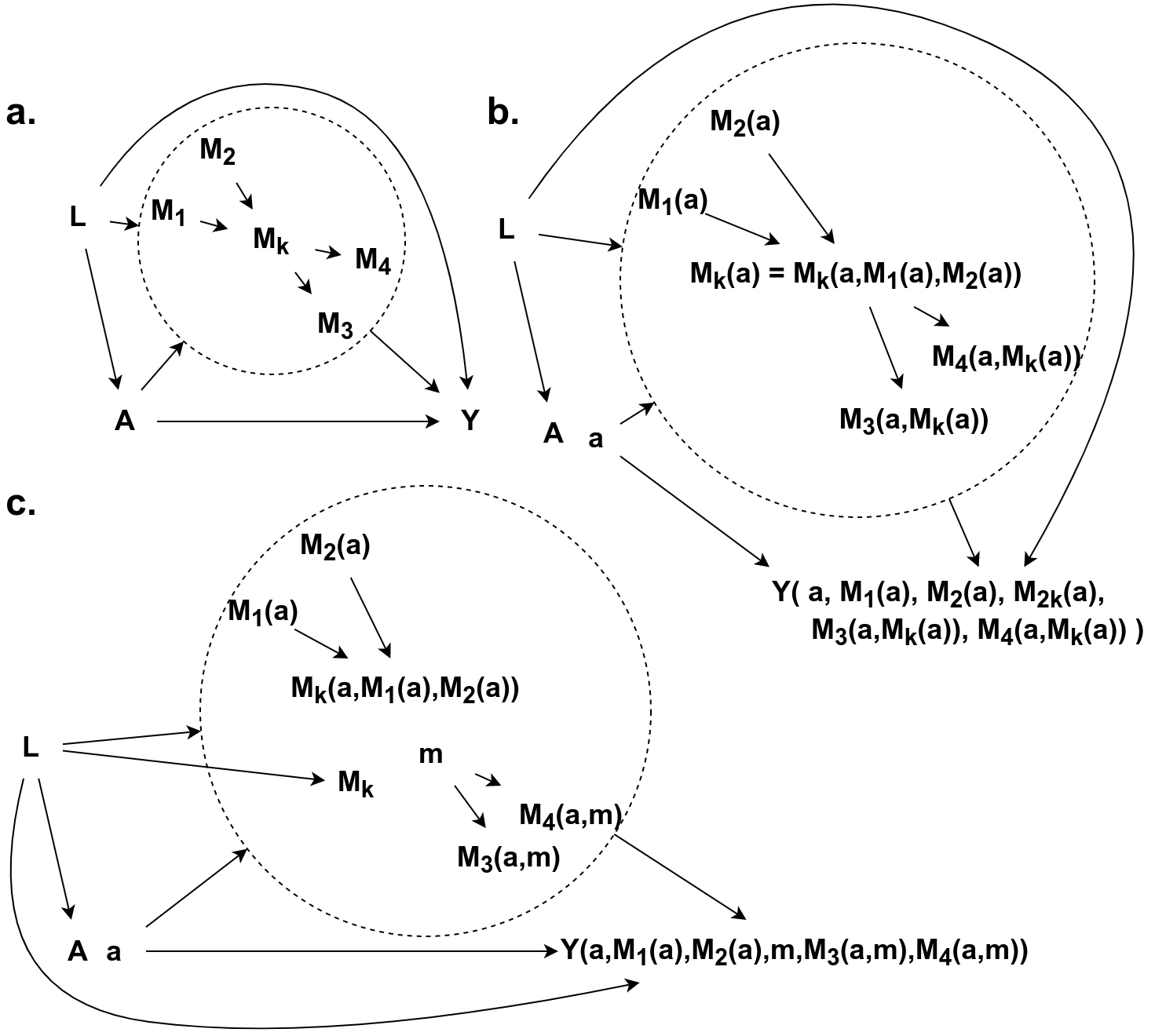}
\centering
\caption{(a) A general causal graph where the mediators in the dashed circle represent multiple mediators with arbitrary causal dependence. Both $L$ and $A$ causally affect each mediator; each mediator causally affect the outcome $Y$. Here we study the $k$-th mediator $M_k$, which has $M_1$ and $M_2$ as its parents and $M_3$ and $M_4$ as its children. (b) The SWIG of panel a when intervening $A$ to $a$, so that the exposure value $a$ and the observed $A$ are separated; and the mediators becomes potential outcome for $a$. We ignored the arrows pointing into the outcome. (c) The SWIG of panel a when intervening $A$ to $a$ and $M_k$ to $k$. Note that there are three versions of $M_k$: $M_k$ is the observed value when no intervention is applied; $M_k(a,M_1(a),M_2(a))$ is the potential outcome of $M_k$ when intervening $A$ to $a$; and $m$ is the intervened value of $M_k$.}
\label{fig:app}
\end{figure}

We can graphically prove Equation~\eqref{eq:ignorability_M} $M_k(a)\ind A\,|\,L$ by constructing the single world intervention graph (SWIG) as in \figurename~\ref{fig:app}b. The conditional independence is true since all connections between $M_k(a)$ and $A$ must go through $L$, which is blocked by conditioning on $L$ based on d-separation.

We can also graphically prove Equation~\eqref{eq:ignorability_Y} $Y_k(a,m)\ind A,M_k\,|\,L$ by constructing the SWIG as in \figurename~\ref{fig:app}c. The conditional independence is true since all connections between $Y_k(a,m)$ and $A,M_k$ must go through $L$, which is blocked by conditioning on $L$ based on d-separation.

\end{document}